\documentclass[preprint,nofootinbib]{revtex4}

\input psfig.sty 
\begin{document}

\bibliographystyle{apsrev}

\title{Dark Energy and Stabilization of Extra Dimensions}

\author{Brian R. Greene${}^{*\sharp}$ and Janna Levin$^{*\dagger}$}
\affiliation{${}^{*}$ Institute for Strings, Cosmology, and
Astroparticle Physics, Columbia University, New York, NY 10027}
\affiliation{${}{\sharp}$ Departments of Physics and Mathematics,
Columbia University, New York, NY 10027} \affiliation{${}^{\dagger}$
Barnard College, Department of Physics and Astronomy, 3009 Broadway,
New York, NY 10027}


\widetext

\begin{abstract}

We discuss the role Casimir energies may play in addressing issues of
moduli stabilization and dark energy. In particular, we examine a
(non-supersymmetric) brane world scenario with toroidal extra
dimensions in which Casimir energies of bulk fields generate a
stabilizing potential for the toroidal volume while driving
accelerated expansion in the non-compact directions. We speculate that
such a scenario might establish a link between asymmetric topology and
asymmetric geometry; that is, asymmetric topology could be linked to
the hierarchy between large and small dimensions.

\end{abstract}

\maketitle



\section{Introduction}

Extra dimensions were proposed almost a century ago in an attempt to
unify gravity with electromagnetism \cite{{kk},{gunnar}}. Although that
particular model of unification has fallen away, the proposal did
raise a simple question: Why are precisely three spatial dimensions
observably large?  Modern attempts at unification, notably
supergravity and string theory, revived the idea of extra dimensions in
a more promising context, but have left the question of ``why three?"
unanswered. What's more, these theories, together with recent
astronomical measurements, have highlighted significant subtleties
such as the need to stabilize the size of the extra dimensions and to
drive periods of accelerated expansion of the large dimensions.

One approach to these issues is to consider them part of the larger
problem of string/M-theory vacuum degeneracy, and invoke anthropically
motivated arguments. However, a more satisfying and convincing
solution would be to find, for example, a dynamical mechanism that
ensures three spatial dimensions grow while the others remain
unobservably small. There have been a number of attempts along these
lines within string/M-theory \cite{{bg},{wb}} but to date these
proposals have fallen short of the mark.

We suggest here that Casimir energies may play an important
role in addressing these questions. Related observations
have been pursued in various guises in earlier works
\cite{{appel},{cw},{chen},{milton},{gupta},{glenn},{acf}}. As is well
known, whenever 
there are dimensions whose spatial 
extent is finite, there are contributions to a purely quantum
mechanical energy density \cite{jaffe} that results from the boundary
conditions 
imposed on a finite space. We argue that if this Casimir energy has
certain features, then it is possible, at least in toy models, to (1)
stabilize the extra dimensions, (2) allow three dimensions to grow
large, and (3) provide an effective dark energy in the large
dimensions. 

It is also intriguing that the lightest field fluctuating
in the compact space will set the size of the dark energy and the size
of the extra dimensions. We would like to draw a connection between
the scale of this dark energy and a neutrino mass and thereby
reduce the number of small parameters in need of explanation. Although our toy
model does not manage to forge this connection (the ``neutrino'' we
will use is not the standard model neutrino for reasons discussed
below), we will mention   
possible scenarios that could.

By way of brief summary, note that the Casimir energies arising from
field fluctuations in large dimensions are insignificant since the
contributions are inversely proportional to some power of the size of
the space. For small extra dimensions, on the other hand, Casimir
energies can be correspondingly large. In fact, by balancing the
Casimir energy contributions of fields with different masses and
spins, the total Casimir energy, as a function of the radius of the
extra dimensions, can develop a non-trivial minimum that stabilizes
their size \footnote{In a different context, this observation was
utilized in \cite{nimaco}. See also \cite{gqr}.}.  And with the size
of the extra 
dimensions fixed, the large dimensions feel ever-more of this Casimir
energy density as they expand, which is the hallmark behavior of dark
energy.
\footnote{One could invoke an additional
mechanism to stabilize the compact dimensions. However, any energy
source capable of stabilizing the extra
dimension, such as a potential for $b$ due to fluxes in string theory,
may well dominate the energy density of the universe and drive inflation on its
own, in which case the
Casimir energy is not providing any fundamentally new physics,
although it will alter the details by lifting the field out of the
minimum of the potential or by changing it's overall shape in a
possibly important way.}

As we discuss below, there are substantial difficulties to realizing a
phenomenologically viable version of this scenario, but it is
intriguing to note that Casimir energies have the potential of
fruitfully linking the issues of moduli stabilization, large/small
dimensions, and dark energy.

\section{Asymmetric Cosmologies}
\label{asym}

There are many ways in which familiar features of isotropic
cosmologies differ significantly when considered in anisotropic
settings. We briefly review two examples that play a key role in what
follows.

First is the famous conclusion that accelerated expansion arises from
a cosmological source with a sufficiently negative pressure ($p_a< -\rho/3$):
\begin{equation}
\frac{\ddot a}{a}=-\frac{4\pi
  G}{3}(\rho+3p_a)\quad\quad ,
\label{uno}
\end{equation}
As this familiar equation assumes isotropy, it is illuminating to consider
an anisotropic spacetime with metric
$ds^2=-dt^2+a_i^2 dx_i^2$. Then, from the Einstein equations we can
make the following comparison: 
\begin{equation}
\dot{H_i}-\dot H_k + \frac{\dot V}{V}(H_i-H_k)=8\pi G(p_i-p_k) \quad
\quad ,\label{differential}
\end{equation}
where the volume is $V=\prod_{i}a_i$ and the $H_i=\dot a_i/a_i$ are the Hubble
factors. For a growing overall volume, the $\dot V/V$ is a
friction term and the pressure differential is a driving force.
In contrast to our intuition from the isotropic case, we can
see that the role of negative pressure is more nuanced
than one would expect. Namely,
negative pressure $p_i$ in the
$i^{th}$ direction acts to \emph{decelerate} expansion in that 
direction relative to the others, while negative
pressure in directions transverse to the $i^{th}$ dimension act to
accelerate expansion relative to the other dimensions. 
In an isotropic
cosmology, all the pressures have equal magnitudes and, as can be seen
from eqn.\ (\ref{uno}), combine to yield a net acceleration in each spatial
dimension. But, fundamentally speaking, negative pressure does not
necessarily entail acceleration.

Second, consider the nature of Casimir energy in an $N+1$ dimensional
spacetime with all of the spatial sections compact with characteristic
size  $b$. The associated stress tensor, $T_{\mu\nu}$ takes the form
\begin{equation}
\left <T_{\mu \nu}\right >=diag(-\rho,\vec p)
\end{equation}
where the energy density has the form $\rho\propto b^{-(N+1)}$, and
the pressure in each dimension, defined as $p=-\partial(\rho
V)/\partial V$ where $V\propto b^N$ is the volume,
is $p_b=\rho/N$. If we now relax
isotropy, say, by taking $n$ of the spatial dimensions to be an
isotropic torus whose radii are set by $b$, while the remaining $N-n$
spatial dimensions comprise ${\bf R}^{N-n}$ (or at least are very
large) and evolve according to the scale $a$, then
\begin{equation}
\left <T_{\mu \nu}\right >=diag(-\rho,\vec p_a , \vec
p_b)
\end{equation}
where, as before, $\rho\propto b^{-(N+1)}$ but now the pressure in the
large directions is $p_a=-\partial(\rho
V_a)/\partial V_a=-\rho$ where $V_a\propto a^{N-n}$. The equation of
state $p_a=-\rho$ follows since $\rho$ does not depend on $a$ and will
hold in the large directions regardless of the topology on the compact space.
By contrast, the pressure in the small directions
is $p_b=-\partial(\rho V_b)/\partial V_b=(N-n+1)\rho/n$, where
$V_b\propto b^n$.  Consistent with
our first observation, it is straightforward to see that this form of
the stress energy tensor gives a negative contribution of
$-(N+1)\rho/n$ to the right hand side in eqn.\ (\ref{differential}),
despite the 
negative pressure in the large
directions. Furthermore, the extra dimensions will expand or contract in
response to the internal pressure so that $H_b$ cannot be
neglected in analyzing the dynamics. (Shear from contracting
directions can even drive inflation 
in vacuum \cite{jl1}.)

In the same spirit of examining features of less symmetric
cosmologies, it is worthwhile to consider Casimir contributions of not
only massless fields (implicit in the expressions above) but also
massive fields. And as we show below, the Casimir energy from massive
fields in a $(3+n+1)$-dimensional universe can lead to an equation of
state that drives a de Sitter epoch in $3$ directions while
simultaneously stabilizing the additional $n$ compact dimensions.

In particular, we show on general grounds that this dynamics is
achieved with an equation of state $p_a=-\rho$ and
$p_b=-2\rho$.  If the Casimir energy is due to a massless
field in a toroidal compactification in $n$ flat directions, then
$\rho\propto b^{-(N+1)}$ and $p_b=(N-n+1)\rho/n$ as noted. However, we show
that if there are massive fields and/or there are other scales at play
-- curvature, warp factors, branes, specific topologies -- then
$p_b=-2\rho$ becomes possible.

While this might seem like a strange equation of state, it is not
pathological as can be confirmed, for example, by looking at an
effective field theory description in which we integrate over the
compact directions. This folds the size of the space $b$ into a radion
field $\Psi$ in an Einstein frame and the Casimir energy into a
potential $U(\Psi)$. As we will see in \S \ref{EF}, the
anisotropic equation of state with $p_b=-2\rho$ is equivalent to a
stable minimum of the potential $U(\Psi)$. The potential, it should be
emphasized, is solely due to the Casimir effect. There is no $\Lambda$,
nor any other effects, added.

A sense of the size of $b$ can be gained by setting the dimensionally
reduced Casimir energy (i.e. the 4-dimensional energy density
$\rho^{(4)}$) equal to the current cosmological constant.  By
dimensional analysis, the dimensionally reduced Casimir energy from a
massless field -- we will later consider massive modifications -- in a
flat spacetime must be of the form
\begin{equation}
\rho^{(4)} \propto b^{-4} \label{rhosimp}
\end{equation}
since $b$ is the only scale in the problem. Given the measured value
of the dark energy \cite{wmap}, a comparison gives
\begin{equation}
  \frac{\alpha}{b^{4}} =
\rho^{(4)}_{DE}\sim (2.3\times 10^{-3}{\rm eV})^4  \quad\quad ,
\end{equation}
which yields a scale of
$b   \sim  {\cal O}\left (  10^{-5} m \right )$
for $\alpha \sim 0.1$, or, in terms of a mass
$b\sim 1/({\rm few}\times 10^{-3}{\rm eV})$. A well known approach for making
such large extra dimensions phenomenologically viable is to consider
standard model fields to be confined to a brane and only gravitational
fields to propogate in the bulk
\cite{{randall},{add},{aadd}}.
It is interesting that $b\sim {\rm few}
\times 10^{-5} m$
is just below the experimental bounds on deviations of
Newton's law \cite{milton}.
Moreover, the mass scale
associated with $b^{-1}$ is in the range of recent neutrino mass
measurements, an observation which will motivate the appearance of
a small scale in our analysis below but one which, as will become clear,
we've yet to incorporate in a fully realistic model.

Before moving on to the details, we note that nothing in our
discussion singles out $3$ for the number
of large, non-compact spatial dimensions. Instead, the point is that
should a mechanism establish a sufficient asymmetry between large and
small dimensions, the effects we describe here can maintain--and in
fact augment--that asymmetry.  Moreover, note too that because Casimir
energies are sensitive to global topology, not every choice of
topology will stabilize the compact dimensions. This leads to the
intriguing possibility that if all of the spatial dimensions are
compact \cite{top}, the reason some expand and others stay small may
be due to the topological form of the large vs.\ small
dimensions. Whatever physical law selects topology may thus have
inadvertently fated three dimensions to grow large.

\section{Cosmological Casimir Dynamics}

We begin with the action for
general relativity in $(3+n+1)$ spacetime dimensions,
\begin{equation}
S=\int d^{4+n}x\sqrt{-g}\left (\frac{M^{2+n}}{16\pi}{\cal R} \right )\quad\quad ,
\label{action}
\end{equation}
(with ${\cal G}=M^{-(2+n)}$)
coupled to a source action whose explicit form we will specify shortly.
We take the metric to be
homogeneous but anisotropic,
    \begin{equation}
      ds^2=-dt^2+a^2(t)d\vec x^2+b^2(t) d\vec y^2
    \end{equation}
where $a$ is the scale factor of the $3$ large directions and $b$ is
the scale factor of the $n$ small directions, and we allow for the
possibility that the large dimensions are finite but large so that any
Casimir effect due to those dimensions is negligibly small. We take
the small dimensions to be compactified on an $n$-torus for
simplicity. In the future it should be interesting to consider the
impact of other topologies.

The Einstein equations can be written as
    \begin{eqnarray}
      3H_a^2+\frac{n}{2}(n-1)H_b^2+3nH_aH_b &=& 8\pi{\cal G} \rho
     \label{1} \\ \dot H_a+3 H_a^2+ n H_aH_b &=& \frac{8\pi{\cal
     G}}{(2+n)}\left [\rho+ (n-1)p_a - n p_b\right ]\label{2} \\ \dot
     H_b+n H_b^2+ 3 H_a H_b &=& \frac{8\pi{\cal G}}{(2+n)}\left [\rho+
     2p_b - 3p_a\right ]\label{3}
    \end{eqnarray}
and the conservation of energy equation is
    \begin{equation}
      \dot\rho+3H_a(\rho+p_a)+nH_b(\rho+p_b)=0 \quad\quad . \label{4}
    \end{equation}
Using eqn.\ (\ref{1}) to eliminate $3H_a^2$ in eqn.\ (\ref{2}) and
requiring that $\dot H_a = 0$ when $H_b,\dot H_b=0$ gives the two
conditions:
\begin{eqnarray}
  -(n+1)+(n-1)w_a-n w_b & = & 0 \label{cond1}\\ 1+2w_b - 3w_a &=& 0
\label{cond2} \quad\quad ,
\end{eqnarray}
where $p_a=w_a \rho$ and $p_b=w_b \rho$.  Using (\ref{cond2}) in
(\ref{cond1}), the extra dimensions are constant and the large
dimensions are inflating if $w_a=-1$ and $w_b=-2$. (We could relax
this condition and require only that $\dot H_a+H_a^2=\ddot a/a >0$ to
get a weaker condition on $w_a$. However, we'll see in the next
paragraph that $w_a=-1$ is also ensured by conservation of energy.)
For an arbitrary number of large dimensions $n_a>1$, the conditions
$\dot H_a=\dot H_b=H_b=0$ give $w_a=-1$ and $w_b=-(n_a+1)/(n_a-1)$. It
is interesting to note that $w_b$ is independent of the number of small
dimensions and that it approaches $-1$ as the number of large
dimensions gets big.

For the extra dimensions to be stably constant requires an additional
condition. The right hand side of eqn.\ (\ref{3}), looks like the
negative of the slope of an effective potential for $H_b$. Stability
requires that effective potential to be concave up. In other words,
the derivative of the right hand side of eqn.\ (\ref{3}) with respect
to $b$ needs to be negative for stability.

The full quantum energy momentum tensor for the Casimir effect from a
massless field can be calculated from the $n$-dimensional Green's
function using a method of images technique to sum over the infinite
copies in the compact space. But, since $b$ is the only scale in the
problem, we can also see dimensionally that $\rho\propto
b^{-(n+4)}$. If $\rho$ is to be independent of $a$, the conservation
equation (\ref{4}) requires the second term vanish and this requires
$w_a=-1$, as did conditions (\ref{cond1}) and (\ref{cond2}). (This
also follows directly from $p_a = -\frac{\partial \left (\rho
V_a\right )}{\partial V_a}$ , and the requirement that
$\rho$ is independent of $a$.)  The conservation equation then reduces
to $\dot \rho=-nH_b(\rho+p_b)$, which can be reexpressed as
\begin{equation}
p_b=-\frac{\partial \left (\rho V_b\right )}{\partial V_b} \quad \quad
,
\label{pbuse}
\end{equation}
where $V_b=b^n$.  For a massless
field then $w_b=4/n$. (In the case of $n=1$, $w_b=N=4$, where $N$ is
the total number of spatial dimensions, as discussed in \S \ref{asym}.) This
equation of state does not stabilize the dimensions nor does it act as
a dark energy.

However, suppose we consider an energy density of the form
\begin{equation}
\rho=\frac{\alpha}{b^{4+n}}\left (1-\beta b^2+\gamma b^4\right ) \quad
\quad .
\label{rhotry}
\end{equation}
This, in fact, is the basic form of the Casimir energy for a light
field ($mb<<1$) with one flat compact direction with periodic boundary
conditions \cite{{wolfram},{nimaco}}.  For scalar fields $\alpha<0$
and for fermions $\alpha >0$.  We will show below that the form
(\ref{rhotry}) results from the sum of Casimir contributions from a
spectrum of massive particles that, for definiteness, we can imagine
to be neutrino-like species.  Other mechanisms to generate a $\rho$
with the properties needed could involve fields from a hidden sector
of string theory with masses of specific relative magnitudes, and/or
certain conditions on the topology. Since a supersymmetric theory
would have equal and opposite contributions from fermions and bosons,
the total Casimir energy would vanish, so $\rho$ also depends on the
specific manner in which supersymmetry is broken. If, for example,
supersymmetry is only broken on a brane \cite{chen}, then the
boundaries set up by a configuration of branes will entirely shape the
Casimir energy. The existence of branes will also generate a different
metric, which might be difficult to determine. Although we'll return
to this discussion below, we show here that a Casimir energy of this
form can realize the scenario described above.  Hereafter we consider
$\beta,\gamma >0$.

\begin{figure}
\centerline{\psfig{file=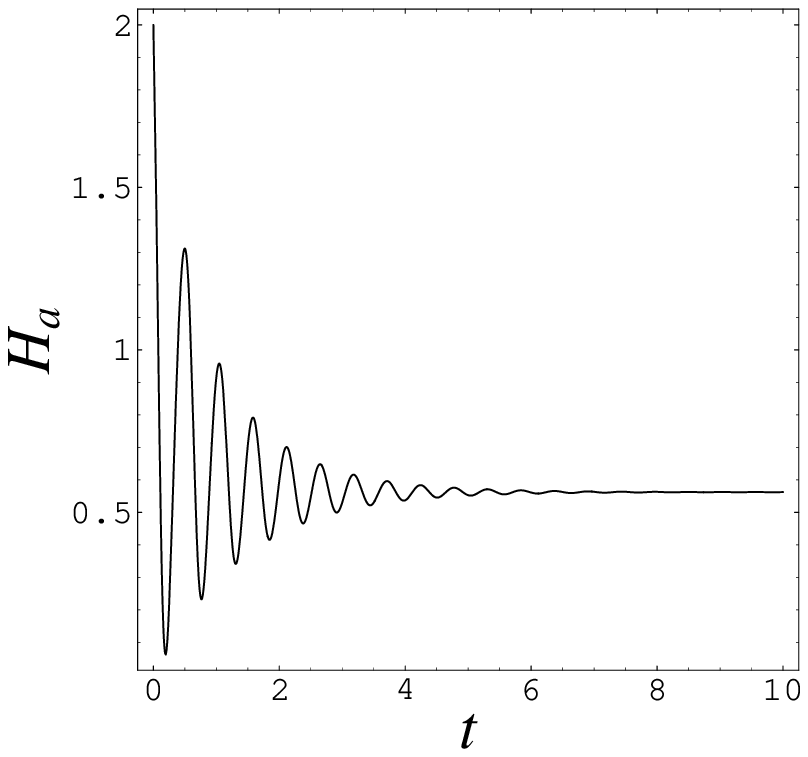,width=2.5in}\quad\quad\quad \psfig{file=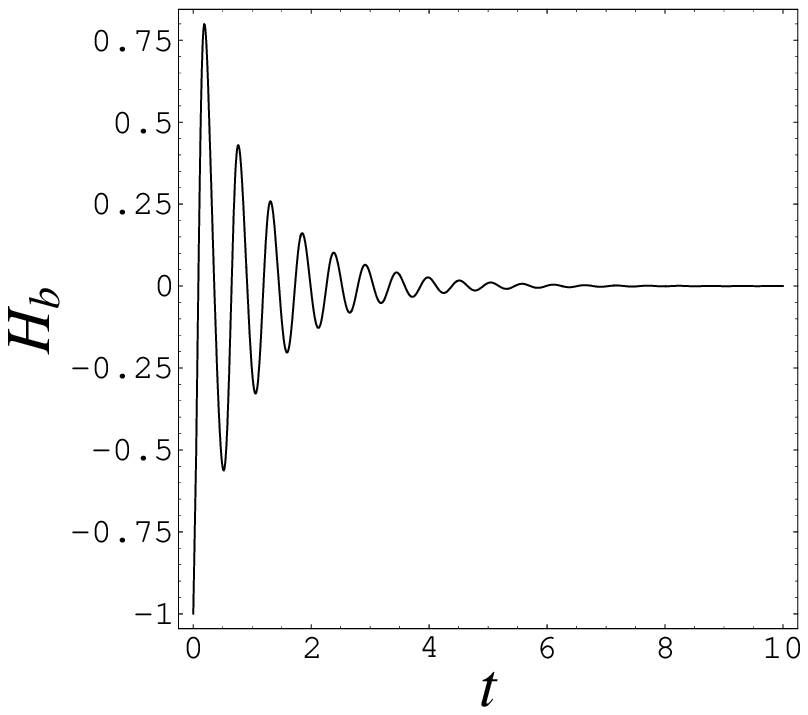,width=2.5in}}
\caption{Left Hand Side: The Hubble constant for the large dimensions,
  $H_a$ as a function of time. Notice the Hubble constant oscillates
  indicating periods of deceleration and acceleration until it settles
  down to a constant value during a de Sitter phase. Right Hand Side:
  The Hubble constant for the compact directions as a function of
  time. $H_b$ oscillates between periods of expansion and contraction
  until it settles down to zero and the dimensions stabilize.
\label{Ht}}  \end{figure}

Using eqn.\ (\ref{pbuse}) gives an expression for $w_b$,
\begin{equation}
w_b = \frac{4}{n}+\frac{2}{n}\frac{\left (\beta b^2-2\gamma b^4\right
  )}{\left (1-\beta b^2+\gamma b^4\right )} \quad\quad .
\end{equation}
The dimensions will be stabilized when $w_b=-2$, which gives the
condition on $b$:
    \begin{equation}
      b_{max,min}^2=\frac{\beta(n+1)\pm
      \sqrt{(n+1)^2\beta^2-4n(n+2)\gamma}}{2n\gamma}\label{extrema}
      \quad \quad .
      \end{equation}
Requiring
\begin{equation}
\frac{\beta^2}{4}<\gamma
<\frac{(n+1)^2}{4n(n+2)}\beta^2\label{condboth} \quad\quad ,
\end{equation}
gives two positive roots (right hand side of (\ref{condboth})) and a
positive energy density at these roots (left hand side of
(\ref{condboth})).  This condition can be satisfied for all $n$
although the window gets pretty narrow as $n$ gets large.

The critical value $b_{min}$ with the negative sign in eqn.\
(\ref{extrema}), corresponds to a stabilization of the extra
dimensions while the critical value $b_{max}$ with the positive sign
in eqn.\ (\ref{extrema}) is an unstable fixed point of eqn.\
(\ref{3}).

\begin{figure}
\centerline{\psfig{file=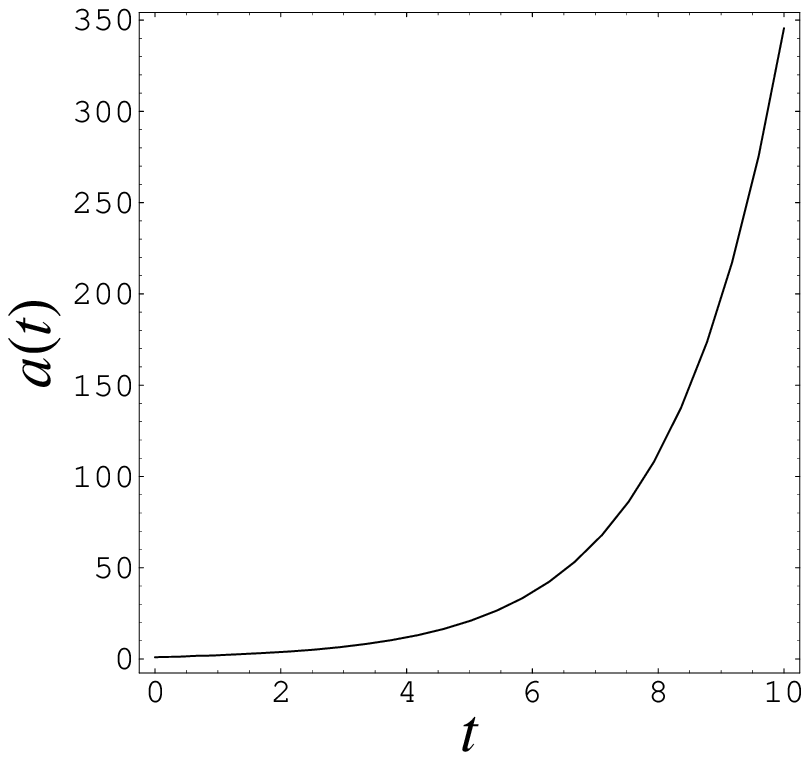,width=2.5in}\quad\quad\quad \psfig{file=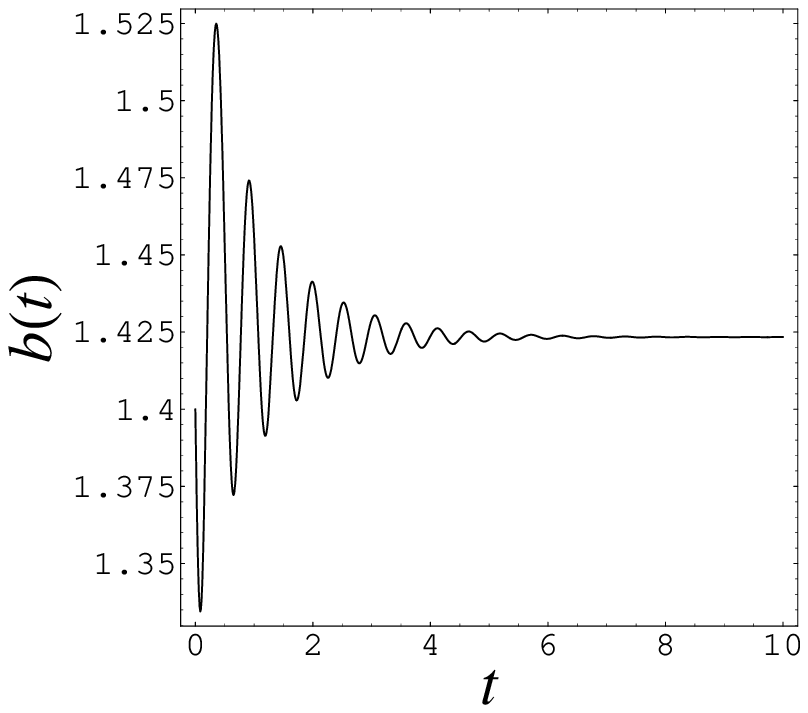,width=2.5in}}
\caption{Left Hand Side: The scale factor for the large directions,
  $a$, as a function of time. Right Hand Side: The scale factor for
  the compact directions, $b$, oscillates about the critical value
  $b_{min}$ until it stabilizes.
\label{rt}}  \end{figure}

Fig.\ \ref{Ht} shows the dynamical evolution of the Hubble factors
under the influence of the energy density (\ref{rhotry}) with $n=2$,
$\alpha=\beta=1$ and $\gamma= 0.9\times((n+1)^2/(4n(n+2)))\beta^2
=0.9\times(9/32)$. The Hubble factor for the large dimensions
alternates in a series of accelerations and decelerations as the scale
factor of the extra dimensions oscillates about the value
$b_{min}$. Eventually $b$ settles at the value $b_{min}$, as shown on
the right hand side of fig.\ \ref{rt}, the dimension stabilizes and
the energy density is constant. The Hubble factor then slides into a
constant value and the large dimensions accelerate as shown on the
left hand side of fig.\ \ref{rt}.

From our perspective in the large dimensions, the universe would
appear to be $(3+1)$-dimensional and to be dark energy dominated. 
To reproduce the strength of gravity today,
$M^{2+n}b^n_{min}=m_p^2$, 
where $m_p$ is
the Planck mass today.

\section{The Radion Picture}
\label{EF}

Since our interest, ultimately, is in how this scenario appears to a
four dimensional observer, it is worthwhile rephrasing our analysis in
the language of four-dimensional effective field theory. Toward this
end, let's augment the action (\ref{action}) by a source term,
$-V(b)$, capturing the Casimir energy contribution discussed in the
last section, and thus start from
\begin{equation}
S=\int d^{4+n}x\sqrt{-g}\left (\frac{M^{2+n}}{16\pi}{\cal R} -
V(b)\right ).\quad\quad
\label{augmentedaction}
\end{equation}
We then integrate over the extra dimensions, and to put the resulting
effective action in canonical form we perform a conformal
transformation to the Einstein frame $g^E_{\mu\nu}= \Omega
g_{\mu \nu}$ ($\mu,\nu=0...3$),
redefine the time variable $dt_E = \Omega^{1/2} dt$, with 
$\Omega=\left (M^{2+n}b^n/{m_p^2} \right )$. Notice that when
$b=b_{min}$, $\Omega=1$.
Under this conformal transformation, the metric is
in standard FRW form with scale factor $a_E(t) = a \Omega^{1/2}$.
Finally,
change field variables to $d\Psi =
\frac{m_p}{\sqrt{16 \pi }} \sqrt{n(n+2)}\ db/b$.
Collectively, these transformations yield
\begin{equation}
S^{\rm {eff}} = \int d^4 x \sqrt{-g_E}\left (\frac{m_p^2}{16 \pi } R[g_E] -
\frac{1}{2}g_E^{\mu \nu} D_{\mu} \Psi D_{\nu} \Psi - U(\Psi)\right )
\end{equation}
from which we derive the equations of motion
 \begin{eqnarray}
H_{E}^{2} &=& \frac{8\pi}{3m_p^2}\left (\frac{1}{2}\left
(\frac{d\Psi}{dt_E}\right )^2 + U(\Psi)\right ) \nonumber \\
\frac{d^2\Psi}{dt_E}+3H_E\frac{d \Psi}{dt_E} &=& -\frac{\partial
U(\Psi)}{\partial \Psi}
    \end{eqnarray}
where $U(\Psi)=Vb^n\Omega^{-2}$ and $H_E=(d a_E/dt_E)/{a_E}$.

\begin{figure}
\centerline{\psfig{file=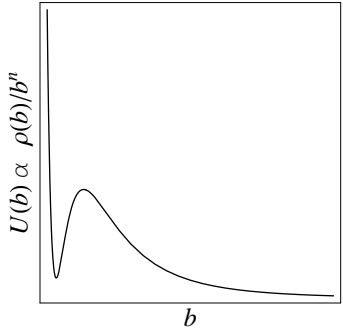,width=1.75in}}
\caption{The potential for the radion field as a function of $b$.
\label{Pb}}  \end{figure}

To make contact with the last section, note that the energy momentum
tensor associated with action (\ref{augmentedaction}) yields the
relations 
	\begin{equation}
{\cal G}\rho = -{\cal G}p_a =  \frac{\Omega}{m_p^2}U\quad\quad ,
\quad \quad
{\cal G}p_b = -\frac{\Omega}{m_p^2} \left
(2U+\frac{b}{n}\frac{\partial U}{\partial b}\right )\quad\quad .\label{relate}
\end{equation}
From this we see that
$p_b=-2\rho$ corresponds to $(\partial U/\partial \Psi)\propto
(\partial U/\partial b) =0 $. In other words, $p_b=-2\rho$ in the
spacetime description corresponds to being at an extremum of the
potential in the radion description. The extrema of $U$ are at
$b_{min,max}$, as must be the case. The potential is drawn as a
function of $b$ in fig.\ \ref{Pb}.

The radion picture is particularly useful for examining forms of
$\rho$ that could stabilize the extra dimensions and create the dark
energy. Previously we mentioned that the sum of the Casimir energies
due to massive fields on a compact torus might have the polynomial
form needed. Here we present a graphical argument for how one can
build a total $\rho$ of the desired form.  In a $(3+n+1)$-dimensional
spacetime with $n$ dimensions compactified to $T^n$, the Casimir
energy density per degree of freedom for massive fields with periodic
boundary conditions \footnote{Different topologies
can lead to boundary conditions other than simply
periodic. Interaction of bulk fields with branes will also change the
modes \cite{{mohammed},{gz1},{gz2}}.} is known to be a sum of Bessel functions
\cite{{wolfram},{szy}}:
\begin{equation}
\rho =\frac{m^{N+1}}{(2\pi)^{(N+1)/2}}
\sum_{j_1=-\infty}^{\infty}...\sum_{j_n=-\infty}^{\infty}\frac{K_{(N+1)/2}(bm\sqrt{j_1^2+...+j_n^2})}{(bm\sqrt{j_1^2+...+j_n^2})^{(N+1)/2}}
\label{bN}
\end{equation}
where $N=3+n$ is the total number of spatial dimensions, $m$ is the
mass of the contributing field. The $j_1=...=j_n=0$ term is infinite
and is subtracted in the usual spirit of renormalizing away the
infinite Minkowski space contribution.  The energy density $\rho$ is
negative for bosons and positive for fermions.

\begin{figure}
\centerline{\psfig{file=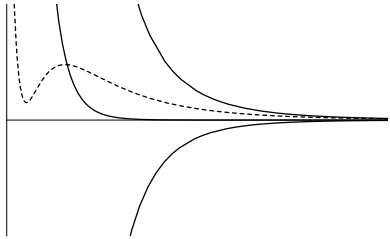,width=2.in}}
\caption{Different contributions to the Casimir energy that sum
  graphically to the desired form. The positive solid curves are
  fermions of different masses while the negative solid curve is a
  scalar contribution of intermediate mass. The dashed line is the
  resultant sum.
\label{scheme}}  \end{figure}

Now, let's consider the Casimir energies due to two fermion fields and
one scalar field, all having different masses (not necessarily light)
as drawn in fig.\ \ref{scheme}.  By adjusting the masses and the
amplitudes through the number of species with a given (similar) mass,
these three graphs can be summed to give a potential with the general
shape of fig.\ \ref{Pb} as indicated by the dotted line in fig.\
\ref{scheme}.

As an illustration, consider a $(4+1)$-dimensional spacetime with one
 direction compactified on a circle of radius $b$. Just for the sake
 of argument consider confining all standard model fields to a flat
 brane as in the ADD model of Ref.\ \cite{{add},{aadd}}. We neglect any brane
 tension as well as any bulk cosmological constant. (Alternatively,
 they can be tuned to cancel.) If the only field propogating in the
 bulk is the gravition, then there is nothing to superpose to create a
 minimum. However, if we posit the existence of sterile fields in the
 bulk (a possibility pursued for other, phenomenologically motivated
 reasons in \cite{{adgm},{cmy}}), we can arrange for a minimum to
 emerge with careful choices of masses and numbers of degrees of
 freedom. We do not justify the existence of these fields
 phenomenologically, as our intention here is simply to demonstrate
 that if such fields are present it is possible to stabilize the
 radion. We limit ourselves to flat internal dimensions but there is
 reason to suppose that curvature or a warp factor could provide the
 additional scale needed to generate a minimum and thereby remove the
 dependence on superposing masses. We leave that for a future
 investigation.

\begin{figure} 
\centerline{\psfig{file=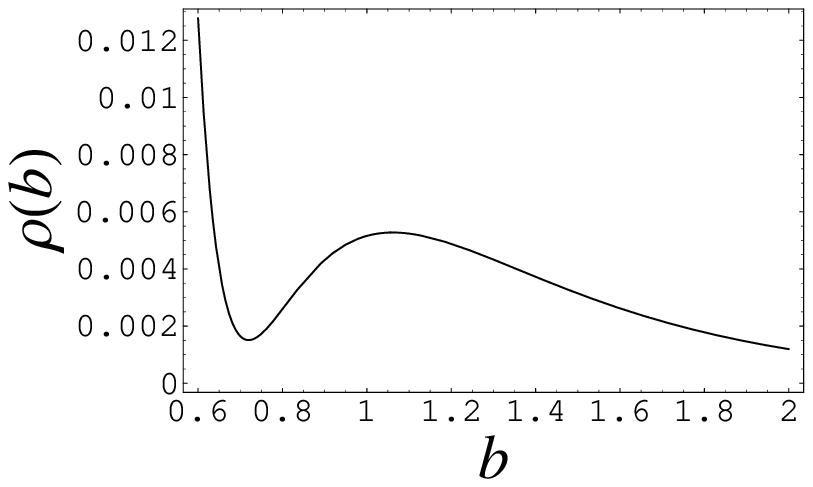,width=2.5in}\quad\quad\quad \psfig{file=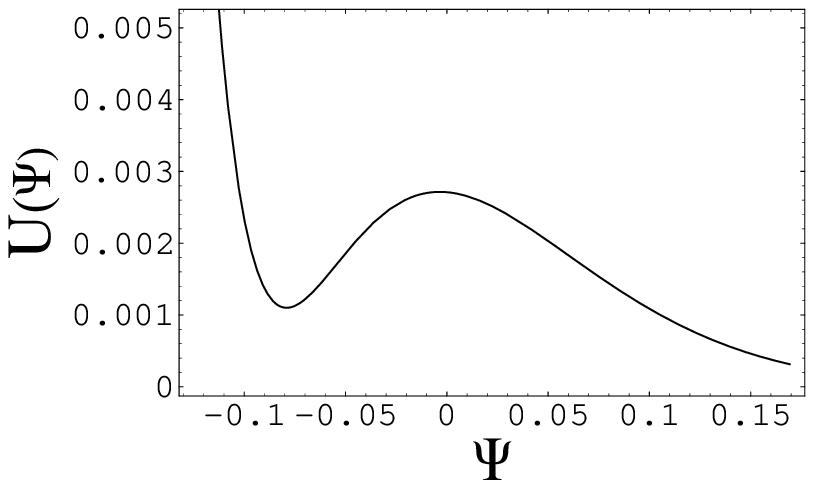,width=2.5in}} 
\caption{Left: The $x$-axis is $b $ in units of $m_{\nu 1}$ and the
  $y$-axis is $\rho$ in units of $m_{\nu 1}^5$. Right: The $x$-axis is
  $\Psi \propto \ln(b\ m_{\nu 1})$ in units of $m_{p}$ and the $y$-axis is
  $U=(b_{min}^2/b(\Psi))\rho(\Psi) $ in units of $m_{\nu 1}^4$.
\label{exists2}}  \end{figure}

We take the particle spectrum in the bulk to consist of 
sterile bulk Majorana neutrinos with masses $m_{\nu
1}\ne 0$, and $m_{\nu 2}=\lambda m_{\nu 1}$ with $\lambda >1$.  In
this toy model we imagine that there are equal and opposite numbers of
massless fermions and bosons. Other combinations of massess and spins
can be used so this is just in way of illustration. We
want $p_b=-2$, which corresponds to a minimum of $U\propto \rho
b^{-1}$. For simply periodic boundary conditions, each term (\ref{bN})
in the sum of the contributions is positive and it is impossible to
add them to get a minimum.  To create a minimum, we add $N_{m_s}=4$
fermion degrees of freedom with anti-periodic boundary conditions and
mass in between the two light neutrinos, $m_s=2m_{\nu 1}$. The Casimir
energy in $(4+1)$ dimensions compactified on $T^1$ with antiperiodic
boundary conditions is \cite{{wolfram},{nimaco}}:
\begin{equation}
\rho =\frac{2m^{5}}{(2\pi)^{5/2}}
\sum_{j_1=1}^{\infty}\frac{K_{5/2}(bmj_1)}{(bmj_1)^{5/2}}\cos(j_1\theta)
\label{b5}
\end{equation}
with $\theta=\pi$.  We can interpret this as the addition of a sterile
Dirac neutrino with mass, $m_s$.  This example establishes that we can
find positive minima, as shown in fig. \ref{exists2}, for a very tight
region, narrower than $3 <\lambda < 3.2$.

Returning to the expansion in terms of $\gamma $ and $\beta $, we can
see why the masses must be so finely tuned. The Bessel functions in
eqns.\ (\ref{bN}) and (\ref{b5}) can be expanded about small $bmj_1$
and summed so that all of the contributions together can be written in
the form
\begin{equation}
\rho =\frac{\alpha }{b^{5}}\left [1-\beta b^2+\gamma b^4\right ]
\quad\quad .
\end{equation}
This approximation is somewhat suspect since the tendency in practice
has been to sum over all $j_1$, long after the condition $bmj_1<<1$
breaks down. We only remark that such an expansion shows clearly that
$\gamma/\beta^2$ is very tightly constrained; Using condition
(\ref{condboth}) for $n=1$,
\begin{equation}
1/4 < \gamma \beta^{-2} < 1/3 \quad \quad .
\end{equation}
This translates directly into the statement that the relative masses
and relative degrees of freedom have to be tuned to create
a positive energy density minimum. For any quantitative results we
continue to use the full Bessel functions.

For comparison with the ADD braneworld scenario proposed to address
the mass hierarchy problem \cite{{add},{aadd}} -- that scenario favored $n=2$
extra dimensions -- we can consider a $(5+1)$-dimensional spacetime
with two directions compactified on a 2-torus of radius $b$ and a
particle spectrum in the bulk consisting of a massless spin 2 boson (2
degrees of freedom) and a massless fermion that we can think of as a
bulk Dirac neutrino (4 degrees of freedom). Also needed is a light
fermion -- another sterile Dirac neutrino (4 degrees of freedom) --
living in the bulk with mass $m_\nu \ne 0$.  We want $p_b=-2$, which
corresponds to a minimum of $U\propto \rho b^{-2}$.  These
fields give a net positive contribution to the Casimir energy.  To get the
negative contribution needed we add $4$ bulk scalar degrees of freedom
with mass $m_s=\lambda m_{\nu}$. (Alternatively, we could interpret
this as another sterile Dirac neutrino with anti-periodic boundary
conditions.)  Numerically, we find positive minima for a very tight
region, narrower than $0.4 <\lambda < 0.42$.

Again, in terms of $\gamma $ and $\beta $, condition (\ref{condboth})
for $n=2$, imposes the restriction
\begin{equation}
1/4 < \gamma \beta^{-2} < 9/32 \quad \quad ,
\end{equation}
or equivalently $0.25 < \gamma \beta^{-2} < 0.28125 $. The range gets
tighter as more additional dimensions are invoked.

\begin{figure}
\centerline{\psfig{file=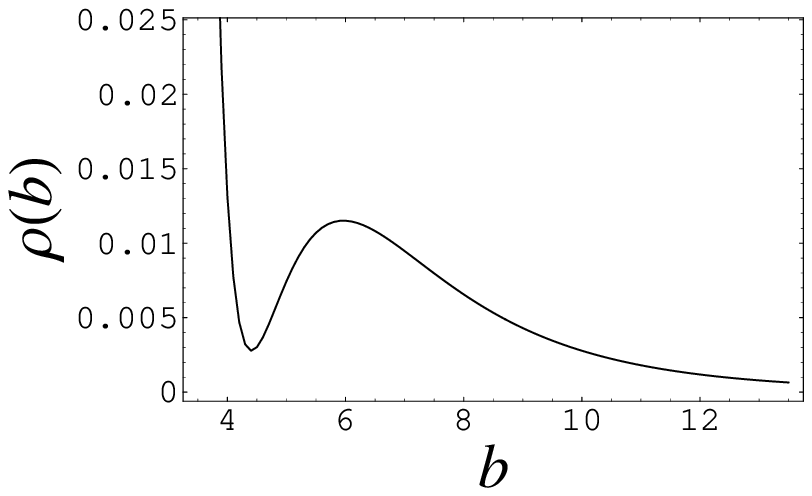,width=2.5in}\quad\quad\quad \psfig{file=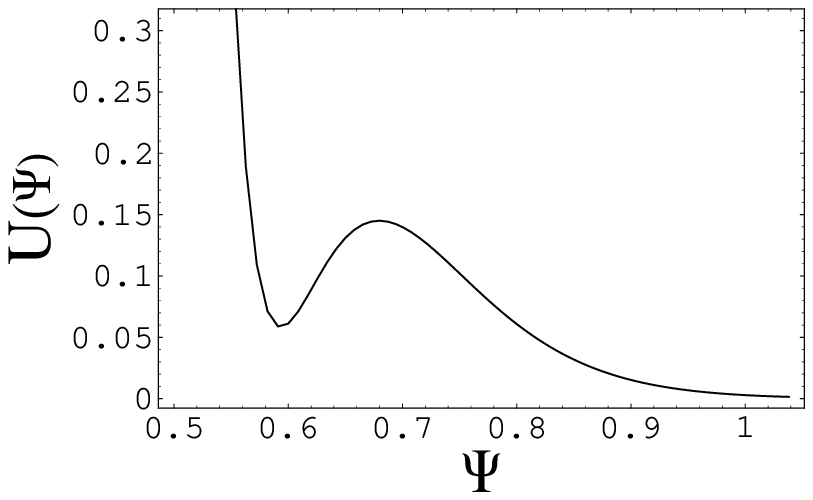,width=2.5in}}
\caption{Left: The $x$-axis is $b $ in units of $m_{\nu }$ and the
  $y$-axis is $\rho$ in units of $m_{\nu }^6$. Right: The $x$-axis is
  $\Psi\propto \ln(b\ m_\nu) $ in units of $m_{p }$ and the $y$-axis is
  $U=(b_{min}^4/b^2)\rho=(b_{min}/b)^2\rho^{(4)} $ in units of $m_{\nu
  }^4$.
\label{exists}}  \end{figure}

For $m_{s}= 0.406 m_{\nu }$, the resultant $\rho $ and $U$ from the
full Bessel functions is shown in fig.\ \ref{exists}.  At the minimum,
$b_{min}\sim 4.5/ m_{\nu }$, the reduced energy density is roughly
$\rho^{(4)}_{DE}=U(b_{min})=\rho b^2_{min} \sim (0.47\times
m_{\nu})^4$. Compare this to the observed value
$\rho^{(4)obs}_{DE}\sim (2.3\times 10^{-3} {\rm eV})^4$. We can choose
$m_{\nu }\sim 5 \times 10^{-3}$eV from which it follows that $b_{min
}\sim 0.2 mm$.  For comparison, the ADD solution to the hierarchy
problem exploits the observation that the Planck mass on the brane is
\begin{equation}
m_{p}^2=M^{2+n}b^n \ \ .
\end{equation}
Turning this around, for $n=2$ and $b=b_{min}\sim 1/(10^{-3}{\rm
  eV})$, the natural scale for $M$ is $\sim 3$ TeV. The size of the
  space needed to fix the dark energy at the observed value is
  therefore consistent with the size needed to address the hierarchy
  problem.  The cosmological constant problem would then be directly
  mappable to the hierarchy problem. Even if the vaccum energy is
  naturally zero, the boundary conditions on the finite extra
  dimensions can create a dark energy of the observed magnitude.

Since the dark energy is really set by the neutrino mass in the bulk,
the picture would feel more complete if such light bulk fields could
also find some justification. It should not go without mention that
the mass $m_\nu \sim 5\times 10^{-3}$eV is a reasonable value for a
neutrino mass and we have been able to express the small dark energy
naturally in these units. The shortcoming is that these are, again,
bulk neutrino fields and not standard model fields.
Thus, from the
perspective of our toy models, the parameters in the effective
Lagrangian are simply chosen to accommodate dark energy. If these ideas
could be incorporated into a more realistic scenario, they might suggest
a mechanism linking neutrino masses and dark energy. Although we've yet
to achieve that link, viable directions to pursue include consideration of
internal manifolds 
whose geometry and topology separates the Casimir and Kaluza-Klein
scales allowing standard model fields to have fluctuations that live
in the bulk. If this
were successful, then the field setting the scale of dark energy could
be a light standard model neutrino thereby reducing the number of
small parameters in need of explanation. 

It is also noteworthy that the
hierarchy between the Planck scale and the electroweak scale
$m_p/M\sim 10^{16}$ is repeated in the hierarchy between the observed
neutrino mass splittings of ${\cal O}(10^{-2}-10^{-3})$ eV and the
electroweak scale: $M/m_\nu\sim 10^{14}-10^{15}$. Attempts to connect
the two have invoked sterile neutrinos in the bulk
\cite{adgm}. However, these models require massless neutrinos in the
bulk and do not offer an explanation for a light massive fermion in
the bulk.

The numerical coincidence between the magnitude of a Casimir term from
macroscopic extra dimensions and the magnitude of the dark energy has
been noted before (see for instance \cite{{milton},{chen},{gupta}}). Here
we're able to generate a proper dark energy with the correct equation
of state ($p_a=- \rho$ and $p_b=-2\rho $, not $p_a=-\rho$, $p_b=0$) so
that the dimensions are fixed and a de Sitter expansion is sourced.
Furthermore, this example hints at a connection between the dark
energy, a light sterile neutrino in the bulk, and the hierarchy
problem.  The ability to sweep up so many seemingly disparate
mysteries into one framework is obviouly appealing. The possibility
that these sterile fields could have heavier Kaluza-Klein excitations
that might be connected with the dark matter adds another layer of
interest to this approach \cite{dm}.

 Nevertheless, it is important to stress that the confining potential
 is generally quite shallow $\Delta U/U \sim 3$ in fig.\ \ref{exists},
 implying that the most minor of perturbations to the energy density
 could unravel the extra dimensions.  The effective mass of the radion
 field, $m_\Psi^2=\frac{\partial^2 U}{\partial \Psi^2}$, would have to
 be big enough to avoid distending the extra dimension
 unacceptably. (For additional discussion of bounds on a light radion
 field in the ADD model, see Ref.\ \cite{adm}.)  A bolder move to
 extend the particle spectrum in the bulk would be required to address
 this problem.  It is also possible that intrinsic curvature in the
 small space could create a deeper, more stable potential.

To demonstrate the destabilizing effect of matter \cite{fr}, consider
the energy 
density in non-relativistic matter today. We allow the matter to live
in the bulk,
\begin{equation}
S=\int d^{4+n}x\sqrt{-g}\left [\frac{M^{2+n}}{16\pi}{\cal R}+{\cal
    L}_{matter}\right ] \quad\quad ,
\label{actionboth}
\end{equation}
although we'll see in the next
paragraph that the effect is ultimately the same if matter is confined
to a brane.
For matter living in the bulk, $\rho_{m}=M/(a^3b^n)$
and $p_M=0$, where $M$ is the total mass of all non-relativistic
matter including baryonic matter as well as any non-relativistic dark
matter. For the sake of argument, suppose all of the dark matter is
cold so that roughly $26$\% of the energy density in the universe is
non-relativistic. The other $74$\% is in dark energy, which we take to
be due to the Casimir energy density today.  In other words,
$\rho_{M0}=(2.6/7.4)\rho_{DE}$.  Using $(a_0/a)=1+z$, we can express
$\rho_M$ as
\begin{equation}
\rho_M=(2.6/7.4)\rho_{DE}^{(4)} (1+z)^3\left (\frac{1}{b^n}\right )
\quad\quad ,
\label{rhom}
\end{equation}
where $\rho_{DE}^{(4)}=\rho_{DE}b_{min}^n$ is the dimensionally
reduced dark energy we measure in $(3+1)$ dimensions today.  The right
hand side of eqn.\ (\ref{3}) becomes $\propto \rho_m+\rho+2p_b-3p_a$
where $\rho$ is still the Casimir energy density and so it is still
the case that $p_a=-\rho$. Stability now requires
\begin{equation}
p_b=-2\rho -\frac{1}{2}\rho_m \label{pbcond}\quad \quad .
\label{pbm}
\end{equation}
Using eqn.\ (\ref{pbm}) in the weaker requirement that $\dot
H_a+H_a^2=\ddot a/a>0$ gives the condition $\rho_m <2\rho
\label{rhomcond}$.  This is the same constraint one gets for a
$(3+1)$-dimensional $\Lambda $ dominated universe. That's to be
expected. When $b$ is stable as required by (\ref{pbm}), the model is
an effective $(3+1)$-dimensional $\Lambda$ dominated universe.

\begin{figure}
\centerline{\psfig{file=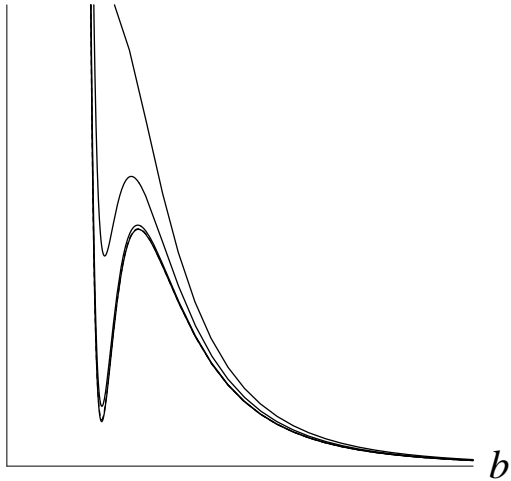,width=2.75in}}
\caption{The potential of figure {\ref{Pb}} with non-relativistic
  matter included. The lowermost black line is without non-relativistic matter
  for comparison. The line that includes non-relativistic matter today
  is dashed but can't be picked out by eye as it
  essentially overlaps the lowermost line. Successive lines as one moves
  up in the figure include
  matter at a redshift of 1,
3.5, and finally 5.5, for which there is no minimum.
\label{matter}}  \end{figure}

In the
radion picture when non-relativistic matter is included, the field is
driven to the minimum of
\begin{equation}
U+\frac{m_p^2{\cal
  G}}{\Omega}\frac{\rho_m}{4} \quad \quad ,
\end{equation}
where $U=(m_p^2{\cal G}/\Omega)\rho$.
Using the dimensionally reduced $\rho_m^{(4)}=\rho_m b^n$, and the
definition of $\Omega$, this can be reexpressed as
\begin{equation}
U+\left (\frac{b_{min}}{b}\right )^{2n}\frac{\rho^{(4)}_m}{4} \quad\quad .
\label{shift}
\end{equation}
Notice that if the non-relativistic matter does not
live in the bulk but is rather confined to a brane,
\begin{equation}
S =\int  d^{4+n}x\sqrt{-g}\left
[\frac{M^{2+n}}{16\pi}{\cal R}\right ]
+ \int d^4x\sqrt{-g^{(4)}}{\cal L}_{matter} \quad\quad ,
\label{actionbrane}
\end{equation}
then there is no
dimensional reduction of the matter action. Under the conformal
transformation, the 4-dimensional matter is shifted,
$\rho_{m}^{(4)}/\Omega^2$, and the radion is again driven to the
minimum of eqn.\ (\ref{shift}).
Today,
the presence of non-relativistic matter would only lift the minimum
slightly.  However, at earlier times, the presence of non-relativistic
matter dominates the shape and eventually destroys the presence of
extrema altogether, as shown in fig.\ \ref{matter}.  The effect is to
force $b$ to expand so that even if a minimum exists today, $b$ has
already sped past it with no way back to its stable position. This may
be curable if there are meta-stable minima to keep $b$ hovering near
its stable value, allowing it to roll gently into today's minimum when
it appears. The point is that it is not enough to understand the shape
of the potential today. The early universe has to be investigated as
well to determine if $b$ can actually make it to its minimum value.

An early universe cosmology that begins supersymmetric would have no
Casimir energy -- except that due to thermal effects. A conformal
anomaly contribution also arises in an expanding space
\cite{bd}. After supersymmetry is broken, any field with $m<1/b$ will
contribute to the Casimir energy and thereby shape the potential for
$b$. As the universe cools, different species contribute to the energy
density in non-relativistic matter leading to an alteration in the
shape of the potential. These issues are currently being investigated
\cite{inprep} to see if the extra dimensions could drive inflation and
then smoothly evolve to generate the dark energy today.

It is perhaps also worth noting that if we are willing to accept some
radical fine tuning, the potential can be made arbitrarily deep. As
$m_s$ decreases to some critical value, the energy density at the
minimum will be pushed closer to zero. Instead of interpreting these
bulk fields as neutrinos, we can suppose they are associated with
electroweak scale physics and have mass $M\sim $TeV. Then $b\sim {\rm
TeV}^{-1}$ and we need not even be on a brane. We could imagine this
in a model of Universal Extra Dimensions (UED) where all standard
model fields propagate in all dimensions \cite{dm}. In units of $M$,
$\rho$ could be arranged to be exceedingly small relative to a TeV,
$\rho\sim (10^{-15}M)^4$ by choosing the relative masses so carefully
that a near cancellation occurs. The potential would be quite stable,
even to the effects of ordinary matter, although one would have to
investigate if tunneling out of the minimum was viable. But, of
course, this extreme fine tunning would be none other than the usual
cosmological constant problem recast. Turning this around, in a model
of UED, Casimir energy contributions to the energy density of the
universe would naturally be huge, with no obvious mechanism for
cancellation -- other than a fortuitous near cancellation between
masses.

\section{Conclusion}

We've shown that Casimir energies--which unavoidably
arise in non-supersymmetric theories with compact extra
dimensions--can act to stabilize the size of the extra dimensions
while also sourcing accelerated expansion in the familiar four
spacetime dimensions. While we've only worked in the context of toy
models, we find it intriguing that there might be a link between
issues of moduli stabilization and dark energy that is a direct
consequence of quantum mechanics on compact spaces.

We've also noted how this approach may link geometrical asymmetries to
topological asymmetries. Namely, because the topology of spacetime's
spatial sections determines the form of the resulting Casimir energy,
different topologies will result in different moduli potentials. Some
potentials will stabilize all but three spatial dimensions, as in the
example discussed herein, while others will stabilize different
numbers of spatial dimensions. In an ensemble of finite
$(3+n+1)$-dimensional universes, the number of dimensions that evolve
to be large or small in a given universe will depend on the
topology. Explaining the existence of $3$ large dimensions may thus
reduce to explaining the topology of the universe.

\bigskip
\bigskip
*Acknowledgement*

We are very grateful for helpful conversations with Lam Hui, Simon
Judes, Amanda Weltman, Pedro Ferreira, and Paul Steinhardt.
BRG
acknowledges financial support from DOE grant DE-FG-02-92ER40699. JL
acknowledges financial support from a Columbia 
University ISE grant.

\end{document}